\documentclass{article}

\PassOptionsToPackage{numbers, compress}{natbib}
\usepackage[position,preprint]{neurips_2026}

\usepackage[utf8]{inputenc}
\usepackage[T1]{fontenc}
\usepackage{hyperref}
\usepackage{xurl}
\usepackage{booktabs}
\usepackage{array}
\usepackage{amsfonts}
\usepackage{amsmath}
\usepackage{amssymb}
\usepackage{nicefrac}
\usepackage{microtype}
\usepackage{xcolor}
\usepackage{graphicx}
\usepackage{multirow}
\usepackage{enumitem}
\usepackage{pgfplots}
\pgfplotsset{compat=1.18}
\title{Position: LLM Inference Should Be Evaluated as Energy-to-Token Production}

\author{%
  \textbf{Xiang Liu}$^{1,*}$ \quad
  \textbf{Shimiao Yuan}$^{2,*}$ \quad
  \textbf{Zhenheng Tang}$^{3}$ \quad
  \textbf{Peijie Dong}$^{1}$ \\[2pt]
  \textbf{Kaiyong Zhao}$^{4}$ \quad
  \textbf{Qiang Wang}$^{5}$ \quad
  \textbf{Bo Li}$^{3,6}$ \quad
  \textbf{Xiaowen Chu}$^{1,\dagger}$ \\[6pt]
  $^{1}$HKUST(GZ) \quad
  $^{2}$UCAS \quad
  $^{3}$HKUST \quad
  $^{4}$XGRIDS \\
  $^{5}$HITSZ \quad
  $^{6}$Guangzhou HKUST Fok Ying Tung Research Institute \\[4pt]
  \small $^{*}$Equal contribution. \quad $^{\dagger}$Corresponding author. \\[2pt]
  \small \href{mailto:xliu886@connect.hkust-gz.edu.cn}{\texttt{xliu886@connect.hkust-gz.edu.cn}} \\[1pt]
  \small Project page: \href{https://dominic789654.github.io/energy-to-token/}{\texttt{dominic789654.github.io/energy-to-token}}
}

\begin{document}

\maketitle

\begin{abstract}
LLM inference is still evaluated mainly as a model or software problem: accuracy, latency, throughput, and hardware utilization. This is incomplete. At deployment scale, the relevant output is a quality-conditioned token produced under joint constraints from effective compute, delivered data-center power, cooling capacity, PUE, and utilization.

We argue that the ML community should treat inference as \emph{energy-to-token production}. We formalize this view with a dimensionally consistent Token Production Function in which token rate is bounded by both compute-per-token and energy-per-token ceilings. Listed API prices vary by over an order of magnitude across providers, but we use price dispersion only as directional motivation, not as causal evidence of marginal cost. The core physical question is instead: under fixed quality and service targets, when does the binding constraint move from theoretical peak compute toward delivered power, cooling, and operational efficiency?

Under this framing, system optimizations---latent KV-cache compression, sparse or heavily compressed attention, quantization, routing, and difficulty-adaptive reasoning---are not merely local engineering tricks. They are energy-to-token levers because they reduce FLOPs/token, joules/token, memory traffic, or utilization losses under fixed $(q^{*},s^{*})$. We therefore call for inference papers and benchmarks to report Joules/token, active binding constraint, PUE-adjusted delivered power, and utilization-adjusted token output alongside accuracy and latency.
\end{abstract}

\section{Introduction}

Tokens are becoming the metered output of AI factories. Each generated token converts electricity, accelerators, memory bandwidth, cooling capacity, and software organization into model output subject to quality and service constraints. This is not a metaphorical analogy. As AI data-center electricity demand rises~\citep{iea2024datacenter,epri2024power} and vendors describe data centers in tokens-per-watt terms~\citep{nvidia2026aifactory,nvidia2026inference}, inference increasingly resembles an industrial production process whose limiting inputs determine both cost and capacity.

Current ML evaluation does not fully reflect this shift. Top-tier inference papers and benchmarks still emphasize accuracy, latency, throughput, and hardware Model FLOPs Utilization (MFU). These metrics remain necessary, but they do not answer the production question: how many quality-conditioned tokens can a deployment produce from a fixed envelope of compute, delivered power, cooling, and utilization? Once that question is asked, system optimizations change meaning. KV-cache compression, sparse attention, quantization, routing, and scheduling are not only micro-level ways to win a benchmark; they are interventions that change the energy-to-token frontier.

Listed LLM API prices make the physical constraint visible, but they do not identify it causally. As of early 2026, posted prices across major providers still span over an order of magnitude on comparable per-million-token units~\citep{openai2026pricing,anthropic2026pricing,deepseek2026pricing}; we use this only as motivation, since the underlying question is whether the binding constraint for generative AI is shifting from theoretical peak compute alone (CapEx) toward delivered data-center power, cooling capacity, PUE, and operational efficiency (OpEx).

This position paper argues that \textbf{LLM inference should be evaluated as energy-to-token production, not merely as model execution.} We formalize this view with a Token Production Function: token output is bounded by both compute-per-token and energy-per-token ceilings under fixed quality and service targets. Under that framing, system optimizations become macro-level energy levers because they reduce FLOPs/token, joules/token, memory traffic, or utilization losses without proportional infrastructure expansion.

Our contribution is fourfold. First, we diagnose why accuracy/MFU-centered inference evaluation is incomplete under regional power and cooling constraints. Second, we formalize quality- and service-conditioned token output with a dimensionally consistent production function. Third, we map concrete inference optimizations onto the physical variables they change: FLOPs/token, Joules/token, memory traffic, and utilization. Fourth, we propose an evaluation agenda: inference papers and benchmarks should report Joules/token, active binding constraint, PUE-adjusted delivered power, and utilization-adjusted token output alongside accuracy and latency.

Our claim is bounded: we do not argue electricity alone determines prices, capability, or geopolitical outcomes, nor treat API prices as causal cost measurements; we argue delivered power and cooling have become binding enough to enter the ML evaluation objective. The paper builds on Green AI~\citep{schwartz2020green}, carbon-accounting work~\citep{patterson2021carbon,patterson2022carbon,wu2023mediation,lacoste2019codecarbon,strubell2019energy,luccioni2024power}, and MLPerf Power~\citep{mlcommons2024mlperfpower}, adding a $(q^{*}, s^{*})$-conditioned Leontief production function, a falsifiable $\rho - \rho^{*}$ diagnostic with a recommended $K_{eff}$ convention, and six disclosure dimensions that turn ``report J/token'' into a comparable benchmark.

\section{The Token Production Function}
\label{sec:production}

To rigorously analyze LLM inference as an industrial process, we propose the following Token Production Function:

\begin{equation}
\dot{Q}_{token}\!\left(t; q^{*}, s^{*}\right)=
\min\!\left(
\frac{K_{eff}(t)}{c_{tok}\!\left(t; q^{*}, s^{*}\right)},
\frac{P_{IT}(t)}{e_{tok}\!\left(t; q^{*}, s^{*}\right)}
\right)\cdot U\!\left(t; q^{*}, s^{*}\right)
\label{eq:production}
\end{equation}

with
\begin{equation}
P_{IT}(t)=\frac{P_{facility}(t)}{PUE(t)},\qquad
Q_{token}=\int_{0}^{T}\dot{Q}_{token}(t; q^{*}, s^{*})\,dt.
\label{eq:production_cumulative}
\end{equation}

This formulation keeps units explicit: $K_{eff}/c_{tok}$ and $P_{IT}/e_{tok}$ are both tokens/sec, and $Q_{token}$ is total tokens over horizon $T$. Importantly, token output is only comparable across systems when evaluated at fixed quality and service targets $(q^{*}, s^{*})$; without this conditioning, token quantity alone is not a meaningful production measure. We define each component:

\begin{itemize}
    \item $Q_{token}$: Total quantity of intelligence tokens produced over time period $T$.
    \item $K_{eff}(t)$: Effective available compute throughput (FLOPs/sec) at time $t$, after hardware availability, kernel efficiency, and memory-stall losses, but before demand-side queueing, batching mismatch, regulatory friction, and operational headroom losses captured by $U$.
    \item $P_{facility}(t)$ and $P_{IT}(t)$: Facility-level power and IT-delivered power (watts), linked by $PUE(t)\ge 1$.
    \item $c_{tok}(t; q^{*}, s^{*})$: Compute intensity (FLOPs/token) at fixed quality target $q^{*}$ and service target $s^{*}$.
    \item $e_{tok}(t; q^{*}, s^{*})$: Energy intensity (joules/token) at the same $q^{*}, s^{*}$ operating point.
    \item $U(t; q^{*}, s^{*})$: Effective utilization factor after the physical ceilings are computed ($0\!<\!U\!\le\!1$), capturing queueing, batching mismatch, request-arrival variability, routing, localization/regulatory friction, and operational headroom.\footnote{$U$ and $\Phi_{system}$ are not literally redundant because they are identified from different signals: $U$ is estimated from \emph{real-time load} (GPU SM activity, queue depth, request arrivals) and captures how much of the deployed capacity is actually in use; $\Phi_{system}$ is estimated from \emph{J/token relative to a physics-limited reference} ($e_{tok}^{ref}/e_{tok}^{obs}$) and captures how much energy the architecture wastes \emph{when fully loaded}. A system can have high $U$ (fully booked) and low $\Phi_{system}$ (architecturally wasteful), or vice versa; the two sources of inefficiency respond to different interventions (provisioning vs.\ algorithmic redesign).}
\end{itemize}

This separation avoids double counting: $K_{eff}$ describes hardware- and execution-level effective throughput, while $U$ describes how much of the resulting physical ceiling is converted into realized token output under demand, scheduling, routing, and institutional frictions. Likewise, $c_{tok}$ and $e_{tok}$ are related but not interchangeable: $c_{tok}$ is computational work demand (FLOPs/token), whereas $e_{tok}$ is measured energy intensity at the operating point (J/token). They therefore define distinct ceilings---compute-throughput capacity and power-delivery capacity---rather than two independent sources of token demand.

The $\min(\cdot,\cdot)$ operator instantiates a Leontief (fixed-proportions) production structure~\citep{leontief1941structure}: compute and delivered power are co-required at a given operating point, not freely substitutable. We adopt it as a local binding-constraint approximation rather than a claim about all long-run technological substitution: it gives the sharpest analytical predictions about which factor is binding when short-run physical substitution is negligible. The CES family~\citep{arrow1961ces} nests both Cobb-Douglas and Leontief as special cases ($\sigma\to0$ gives Leontief); we use Leontief as the binding-constraint limit. Under this form, $\Phi_{system}$ improvements do not substitute one factor for another at a fixed technology---they \emph{shift} the production frontier by simultaneously reducing $c_{tok}$ and $e_{tok}$ (or raising $U$), rescaling both arms of the $\min$ together. This is why Section~\ref{sec:optimizations}'s architectural gains (MLA, NSA, hybrid linear attention) are consistent with a Leontief structure: they relax both the compute and delivered-power constraints, rather than trading FLOPs for joules at a fixed operating point. As data-center power densities exceed 100\,kW/rack~\citep{uptime2024survey}, $P(t)$ has emerged as the scarce factor in many regions.

To avoid over-aggregation, we treat $\Phi_{system}$ as a structured set of mechanisms that parameterize $c_{tok}$ and $e_{tok}$ rather than a single free multiplier:
\[
\Phi_{system}\equiv\{\Phi_{prefill},\Phi_{decode},\Phi_{mem},\Phi_{comp},\Phi_{sched},\Phi_{route}\},
\]
with $c_{tok}=c_{tok}(m,w,\Phi_{system})$ and $e_{tok}=e_{tok}(m,w,\Phi_{system},PUE)$ for model/workload pair $(m,w)$. This decomposition is necessary because some interventions help prefill but not decode, or trade off energy against latency/quality.

\textbf{Operational estimation.} Each $\Phi$ component admits a ratio-form estimator: $\Phi_{mem}\approx\dot{Q}_{obs}/\dot{Q}_{BW}^{ceil}$ with $\dot{Q}_{BW}^{ceil}=BW_{HBM}/(2N_{param}\cdot w_{bytes})$ from hardware specs; $\Phi_{decode}$ restricts the numerator to decode-phase tokens; $\Phi_{sched}\approx\bar{U}_{SM}/U_{SM}^{batch^{*}}$ from SM-activity counters (e.g., \texttt{DCGM\_FI\_PROF\_SM\_ACTIVE}) divided by the ideal-batch reference; aggregate $\Phi_{system}\approx e_{tok}^{ref}/e_{tok}^{obs}$ against a dense MHA at FP16 baseline at the same parameter count~\citep{samsi2023wordstowatts,niu2025tokenpowerbench}. Values $\Phi_{mem}<0.3$ indicate memory-bound operation; $\Phi_{sched}\ll1$ indicates scheduling/batching overhead.

This bridges systems engineering, macroeconomics, and energy policy: $K(t)\leftrightarrow$CapEx, $P(t)\leftrightarrow$OpEx, and $\Phi_{system}\leftrightarrow$TFP in the sense of Solow~\citep{solow1957technical}---the residual output gain from better organization rather than raw input expansion. Unlike a pure macroeconomic residual, however, $\Phi_{system}$ is partially decomposable into measurable serving mechanisms.

\textbf{Which constraint binds?}
The $\min(\cdot,\cdot)$ structure raises a practical question: when is compute the binding factor and when is delivered power? The crossover occurs at the \emph{constraint boundary}:
\begin{equation}
\frac{K_{eff}}{c_{tok}} = \frac{P_{IT}}{e_{tok}}
\quad\Longleftrightarrow\quad
\frac{P_{IT}}{K_{eff}} = \frac{e_{tok}}{c_{tok}} \equiv \rho^{*},
\label{eq:crossover}
\end{equation}
where $\rho^{*}$ (joules/FLOP) is the \emph{energy-per-FLOP ratio demanded by the workload}. If $\rho \equiv P_{IT}/K_{eff} > \rho^{*}$ compute is scarce; if $\rho < \rho^{*}$ delivered power is scarce. Eq.~\ref{eq:crossover} extends the Roofline binding-constraint logic~\citep{williams2009roofline} from memory bandwidth to delivered data-center power, conditioned on $(q^{*}, s^{*})$. The regime classification depends on whether $K_{eff}$ is measured as theoretical peak compute or as realized serving throughput, since memory stalls, insufficient batching, and utilization losses can move the same deployment between regimes. \textbf{We therefore recommend a fixed reporting convention}: $K_{eff}$ should default to \emph{realized effective serving throughput at the disclosed $(q^{*}, s^{*})$ operating point} (with batching, context length, and energy-accounting boundary stated), and peak-throughput $K_{eff}$ may be reported alongside as an upper-bound calibration only. Under this convention $\rho - \rho^{*}$ becomes a falsifiable diagnostic: a deployment whose realized $\rho$ exceeds its workload $\rho^{*}$ at the stated operating point is, by construction, not power-bound. Appendix~\ref{app:rho} works through a 65B-class anchor on H100 to show how the same hardware can be classified as power-bound under a peak-throughput denominator and effective-compute-bound under a realized-throughput denominator. As context lengths grow and KV-cache bandwidth dominates, $c_{tok}$ and $e_{tok}$ shift together with the operating point, and regions with tight grid headroom enter the power-bound regime first. This constraint-switching logic explains why the same model family can appear effective-compute-bound in a well-powered, well-utilized campus and power-bound in a capacity-constrained region.

When delivered power is the bottleneck, improvements that reduce measured $e_{tok}$ expand effective capacity without additional infrastructure: a memory-efficiency gain that cuts J/token by 50\% raises the power-side token ceiling under the same power cap without adding a single watt. \textbf{What counts as a $\Phi_{system}$ gain.} A gain only ``counts'' when it preserves the operating point: retrieval and reasoning quality must remain within disclosed tolerances of the reference (e.g., MMLU within $\epsilon$ and a long-context benchmark such as RULER or IFEval within $\delta$ at the stated context length), latency must stay within the $s^{*}$ envelope, and reliability/freshness must not regress; gains that fail these checks shift the operating point and are not directly comparable. Under these fixed targets, inference papers should report not only accuracy, latency, throughput, and MFU, but also J/token, the active binding constraint, PUE-adjusted delivered power, and utilization-adjusted token output.

\begin{figure}[t]
\centering
\includegraphics[width=0.9\textwidth]{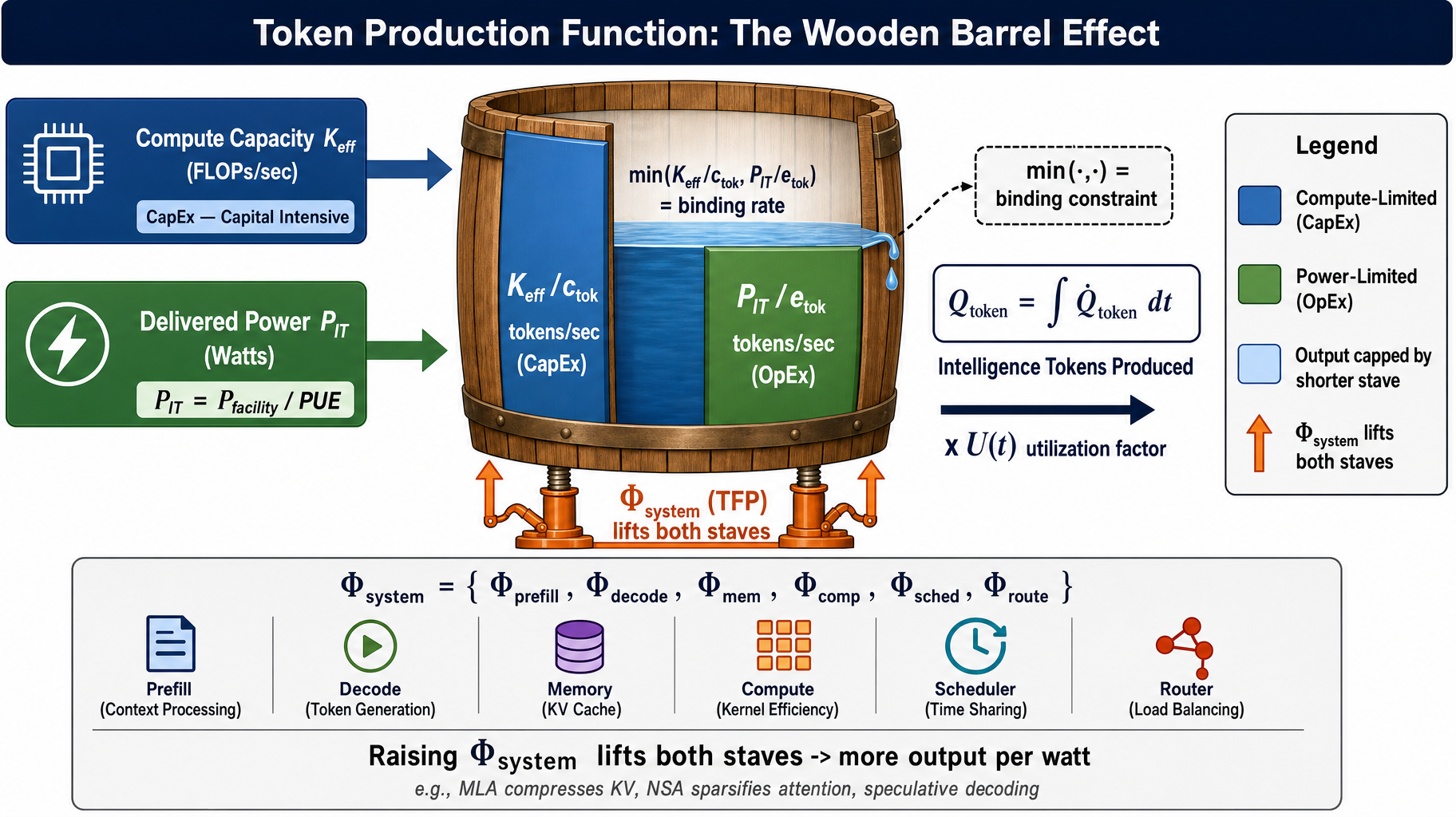}
\caption{The thermodynamics of token generation, illustrating how the Token Production Function converts physical resources (compute $K$ and delivered power $P$) into intelligence tokens through system-level optimizations $\Phi_{system}$. The $\min(K_{eff}/c_{tok}, P_{IT}/e_{tok})$ constraint creates a ``wooden barrel effect'' where the limiting rate determines total output.}
\label{fig:thermodynamics}
\end{figure}

\section{When Power Becomes the Binding Constraint}
\label{sec:validation}

\begin{figure}[t]
\centering
\includegraphics[width=0.97\textwidth]{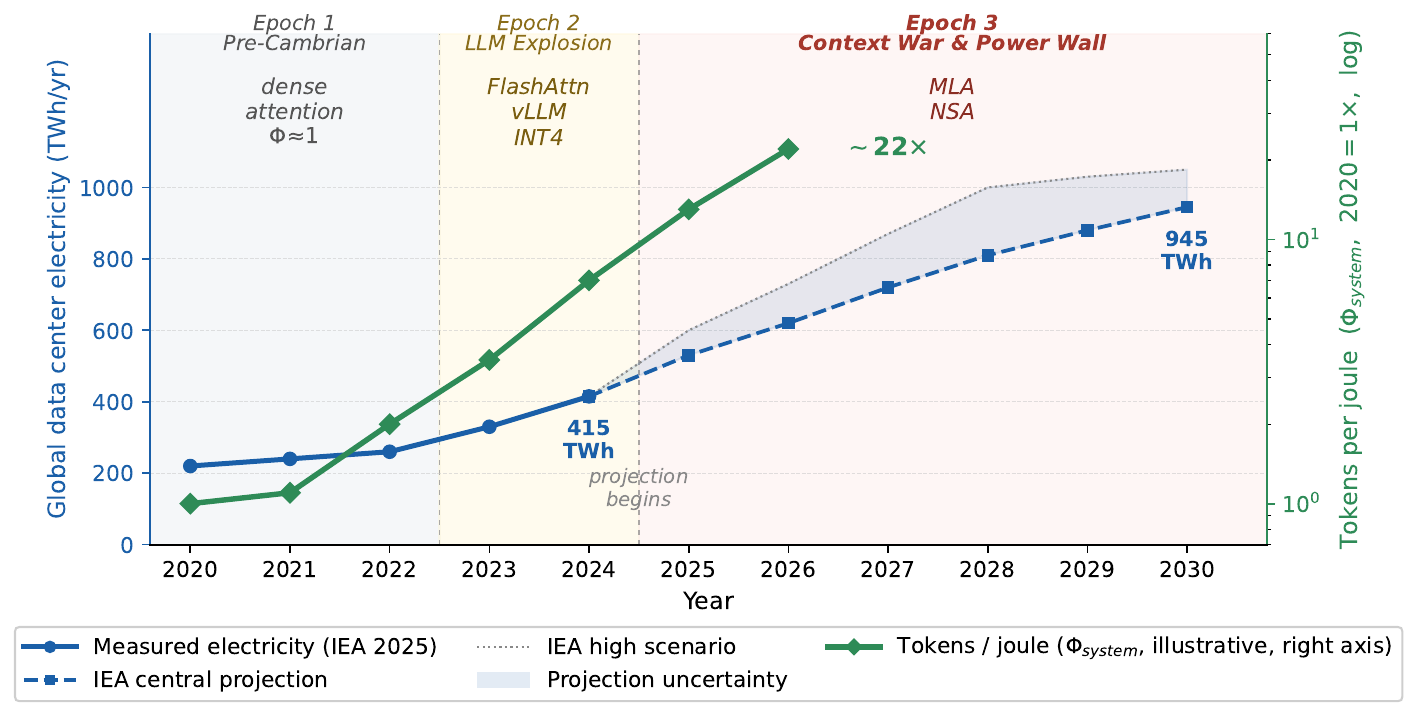}
\caption{\textbf{Left axis (blue):} global data center electricity (TWh/yr), 2020--2030 (IEA measured, central, and high scenarios with projection band)~\citep{iea2024datacenter,iea2025ainews}. \textbf{Right axis (green, log):} illustrative $\Phi_{system}$ proxy normalized to 2020$=$1, with step-changes anchored to documented system-level deployments (Epoch 2: FlashAttention, vLLM/PagedAttention, INT4/AWQ; Epoch 3: MLA, NSA, sparse-hybrid). Energy grows roughly linearly while $\Phi_{system}$ rises over an order of magnitude---tokens partially decouple from joules. The proxy is a qualitative visualization, not a fitted measurement; methodology and caveats are in \S\ref{sec:validation}.}
\vspace{-10pt}
\label{fig:token-trend}
\end{figure}

We use the Token Production Function as an interpretive lens to organize inference history into three epochs. \textbf{Methodological note}: throughout this paper, comparisons between API prices and regions are treated as \emph{directional association}, not causal identification---posted prices are not normalized for quality, latency SLOs, context windows, caching, or subsidy strategies. Similarly, this section is a \emph{theoretical framework illustration}, not an empirical validation: annual proxies for $P_{facility}$ and $K_{eff}$ are mapped to public data~\citep{iea2024datacenter,sevilla2024training}; $\Phi_{system}$ is inferred qualitatively from documented step-changes. No causal claims are made anywhere in the paper unless explicitly stated. Figure~\ref{fig:token-trend} anchors a proxy for average $P_{facility}(t)$ to IEA annual electricity consumption (TWh/yr $\div$ 8760 h/yr; Eq.~\ref{eq:production} uses power, not energy). Epoch boundaries mark $\Phi_{system}$ step-changes that partially decoupled token output from energy growth. Table~\ref{tab:anchors} gives order-of-magnitude calibration anchors~\citep{niu2025tokenpowerbench,chung2026joules,delavande2026efficiency,cavagna2026sweetspot}.

\begin{table}[t]
\centering
\footnotesize
\caption{Order-of-magnitude anchors for Eq.~\ref{eq:production} variables (2024--2026).}
\vspace{-5pt}
\label{tab:anchors}
\setlength{\tabcolsep}{4pt}
\renewcommand{\arraystretch}{0.92}
\begin{tabular}{>{\raggedright\arraybackslash}p{3.0cm} >{\raggedright\arraybackslash}p{10.0cm}}
\toprule
Variable (unit) & Anchor value (source) \\
\midrule
$K_{eff}$ (FLOPs/s) & H100 peak: TF32 $9.89\times10^{14}$, BF16 $1.979\times10^{15}$; 8$\times$H100 BF16 $\approx1.6\times10^{16}$~\citep{nvidia2026h100,nvidia2026hgx}. \\
$P_{facility}$ (avg. GW) & Global data centers: $\approx47$ GW in 2024 (415 TWh/yr), $\approx108$ GW in 2030 IEA central projection (945 TWh/yr)~\citep{iea2024datacenter,iea2025ainews}. \\
$PUE$ (ratio) & Industry avg. $\approx1.56$; leading sites $\approx1.08$--$1.09$~\citep{uptime2024survey}. \\
$e_{tok}$ (J/token) & 65B-regime anchor $\approx3$--$4$, with large workload/serving variance~\citep{samsi2023wordstowatts,chung2026joules,delavande2026efficiency,luccioni2024power}. \\
$c_{tok}$ (FLOPs/token) & Dense proxy $\sim2\times10^{9}$--$\sim8\times10^{11}$ for 1B--405B models using $c_{tok}\!\approx\!2N$~\citep{niu2025tokenpowerbench}; for autoregressive decode under KV reuse, replace with weight-read + per-step attention FLOPs, and for MoE serving substitute $N_{active}$ for $N$. \\
$U$ & Observed range 0.3--0.7; batching, scheduling, and request mix dominate~\citep{chung2026joules,delavande2026efficiency}. \\
\bottomrule
\end{tabular}
\end{table}

\noindent \textbf{Directional calibration.} Table~\ref{tab:empirical} gathers representative $e_{tok}$ values for 65B-class inference from independent sources; it is an illustrative compilation, not a single controlled head-to-head benchmark. Rows differ in serving stack and workload mix, and the 65B / 100\,ms SLO framing is a nominal anchor rather than a normalized ceteris-paribus comparison. The table's purpose is to show the \emph{direction} and \emph{rough magnitude} of $\Phi_{system}$ effects (architecture and quantization lower $e_{tok}$ without expanding $K_{eff}$ or $P_{facility}$ budgets), which is consistent with---though not a controlled test of---the claim that optimization acts as an energy multiplier.

\begin{table}[t]
\centering
\small
\caption{Representative $e_{tok}$ values for 65B-class LLM inference at a nominal $(q^*, s^*)$ anchor (MMLU/IFEval-class quality, 100\,ms latency). \textbf{Rows A--C are measured} from the cited independent sources under the listed configurations; \textbf{row D is a projection} composing the KV-compression batch headroom of DeepSeek-V2~\citep{deepseek2024mla} with the INT4 energy gains reported by~\citep{delavande2026efficiency}, not a matched-stack measurement. Stack, workload mix, batching, and energy-accounting boundary differ across rows.}
\vspace{-5pt}
\label{tab:empirical}
\begin{tabular}{c >{\raggedright\arraybackslash}p{3.0cm} >{\raggedright\arraybackslash}p{2.8cm} c c >{\raggedright\arraybackslash}p{3.0cm}}
\toprule
& Configuration & Implementation & $e_{tok}$ (J) & Rel.\ MHA & Source \\
\midrule
A & Standard MHA + FP16 & H100, batch=8 & $\approx 3.5$ & $1.0\times$ & measured~\citep{samsi2023wordstowatts,chung2026joules} \\
B & MHA + INT4 quant. & H100, batch=8 & $\approx 1.2$ & $0.34\times$ & measured~\citep{delavande2026efficiency} \\
C & MLA (KV compress.) & H100, batch=24 & $\approx 1.1$ & $0.31\times$ & measured~\citep{deepseek2024mla,niu2025tokenpowerbench} \\
\midrule
D & MLA + INT4 & H100, batch=24 & $\approx 0.35$ & $0.10\times$ & \emph{projection} \\
\bottomrule
\end{tabular}
\end{table}

\noindent The measured A$\to$C spread is $\sim$3$\times$; the additional 3$\times$ implied by composing INT4 onto MLA (row D) is a projection. The framework's conservative claim is that architecture-side $\Phi_{system}$ levers move $P_{IT}/e_{tok}$ by at least the measured 3$\times$, with $\sim$10$\times$ plausible when quantization composes; a controlled cross-stack J/token benchmark closing this gap is what the reporting agenda calls for.

\subsection{Epoch 1 (2020--2022): The Pre-Cambrian Era}

In the early phase, both $K(t)$ and $P(t)$ were abundant relative to demand. GPT-3-scale models ran on concentrated clusters with $\Phi_{system} \approx 1$---no sophisticated memory management or scheduling. The field operated under scaling laws suggesting strong returns from parameters and compute~\citep{kaplan2020scaling, hoffmann2022training, rae2021scaling}; energy costs were buried in operational budgets.

\subsection{Epoch 2 (2023--2024): The LLM Explosion}

ChatGPT triggered exponential $K(t)$ growth~\citep{desislavov2023compute, sevilla2024training} alongside the first wave of $\Phi_{system}$ improvements. FlashAttention~\citep{dao2022flashattention} reduced attention memory movement from $O(N^2)$ to $O(N)$, lowering both $c_{tok}$ and $e_{tok}$; PagedAttention/vLLM~\citep{kwon2023efficient} enabled dynamic KV-cache allocation; INT4/INT8 quantization~\citep{frantar2023gptq, lin2024awq} stretched $K(t)$ within existing $P(t)$ envelopes. Empirical runtime profiling of training, fine-tuning, and inference on commodity hardware confirmed early on that memory traffic, not raw FLOPs, dominates real-world LLM throughput~\citep{zhang2023dissecting}. API pricing remained relatively uniform---energy was not yet the binding constraint.

\subsection{Epoch 3 (2025--2026): The Context War and Power Wall}

Context lengths have reached 1M+ tokens, motivating long-context generation benchmarks~\citep{liu2024longgenbench} for evaluation under sustained-output workloads, and the Power Wall has emerged as a binding constraint. Global data center electricity reached 415\,TWh in 2024 and is projected to reach 945\,TWh by 2030~\citep{iea2024datacenter,iea2025ainews}; US data centers alone may reach 325--580\,TWh by 2028~\citep{epri2024power,doe2024datacenter}. US hyperscaler capex has grown $\sim$72\%/yr since Q2 2023, exceeding \$400\,B in 2025~\citep{bigtechcapex2024}; on the demand side, China reported $\sim$140\,T daily token calls by March 2026 ($\sim$1000$\times$ early 2024; ByteDance Doubao alone $\sim$120\,T/day)~\citep{nda2026liu,technode2026doubao}. Some regions have hit the $P(t)$ ceiling, and the API price divergence is consistent with this constraint divergence.

\section{System Optimizations Are Energy Multipliers}
\label{sec:optimizations}

$\Phi_{system}$ summarizes phase- and mechanism-level choices that can reduce $c_{tok}$ and $e_{tok}$ under fixed quality/SLO and measurement assumptions. We examine two mechanisms through which micro-level engineering decisions can become macroeconomic energy levers, while treating reported speedups and energy reductions as configuration-dependent rather than universal constants.

\subsection{Latent Compression Moves the Memory Boundary}
\label{sec:mla}

KV-cache memory bandwidth is the dominant bottleneck in long-context inference: saturated HBM leaves compute units idle, wasting both CapEx and OpEx~\citep{wulf1995memory}. We use one publicly documented attention lineage to illustrate how memory-side $\Phi_{system}$ levers compose. DeepSeek-V2 introduced Multi-head Latent Attention (MLA)~\citep{deepseek2024mla} for low-rank KV compression, and NSA added learned sparse selection~\citep{deepseek2025nsa}. The DeepSeek-V4 technical report~\citep{deepseekv42026} is one example of a hybrid compression-and-sparsity stack: Compressed Sparse Attention (CSA) compresses KV blocks before top-$k$ selection, Heavily Compressed Attention (HCA) applies more aggressive compression with dense attention over the compressed representation, and these are layered with FP4-trained indexing, multi-head hybrid compression, and heterogeneous KV-cache placement across HBM, CPU memory, and SSD. The report targets 1M-token context serving and lists only $\sim$27\% of V3.2 single-token FLOPs and $\sim$10\% of V3.2 KV cache (developer report, pending third-party replication). Other production stacks combine subsets of the same levers---paged KV management in vLLM~\citep{kwon2023efficient}, FlashAttention IO scheduling~\citep{dao2022flashattention}, eviction-based KV reduction~\citep{zhang2024chunkkv,zhang2023h2o,liu2025flowkv,liu2026semanticintegrity,chen2026sonic,li2025antkv,zhu2025oraclekv}, and offloaded inference~\citep{sheng2023flexgen}---and we cite this lineage as one observed instance, not as the recommended architecture. Compression counts as a production-function gain only if retrieval, reasoning, latency, and reliability remain within the fixed $(q^*,s^*)$ envelope; under that constraint, the family of memory-side optimizations enables:

\begin{enumerate}
    \item \textbf{Higher batch sizes}: more concurrent sequences within the same memory envelope, potentially increasing throughput per watt under comparable latency targets.
    \item \textbf{Million-token contexts}: routinely supporting 1M-token inputs on hardware that would otherwise be memory-bound at far shorter sequence lengths.
    \item \textbf{Improved hardware utilization}: reducing the time compute units spend stalled on memory transfers when memory traffic is the binding bottleneck.
\end{enumerate}

Prior work on semantic-preserving KV cache compression via eviction and offloading reports up to 50\% cache reduction under task-specific quality constraints~\citep{zhang2024chunkkv, zhang2023h2o, sheng2023flexgen}; the DeepSeek lineage extends this with learned compression and sparse top-$k$ selection. These methods compound $\Phi_{mem}$, $\Phi_{comp}$, and $\Phi_{prefill}$ only when the reduced cache preserves task-relevant evidence---compression that degrades retrieval is not a pure efficiency gain. Under comparable measurement assumptions the reported direction is an order-of-magnitude reduction in $e_{tok}$ and $c_{tok}$ at million-token context. Appendix~\ref{app:mla} gives a worked bandwidth derivation.

\noindent\textbf{Cross-vendor price evidence.} As of April 2026, the tier-matched output-price gap between frontier Chinese reasoning Pro tiers (\$1--\$4/M) and frontier US Pro/Sonnet tiers (\$12--\$30/M) is roughly \textbf{5--10$\times$}; the wider 3--30$\times$ envelope cited in some reports compares Flash-tier Chinese models to frontier US Opus/GPT-5 tiers and is therefore cross-tier, not like-for-like (Appendix~\ref{app:prices} gives the per-vendor breakdown). The gap is consistent with infrastructure-level $\Phi_{system}$ differences shaping marginal economics, alongside quality, latency-SLO, and business-model variation; we do not attribute it causally to any single factor.

\begin{figure}[t]
\centering
\includegraphics[width=1\textwidth]{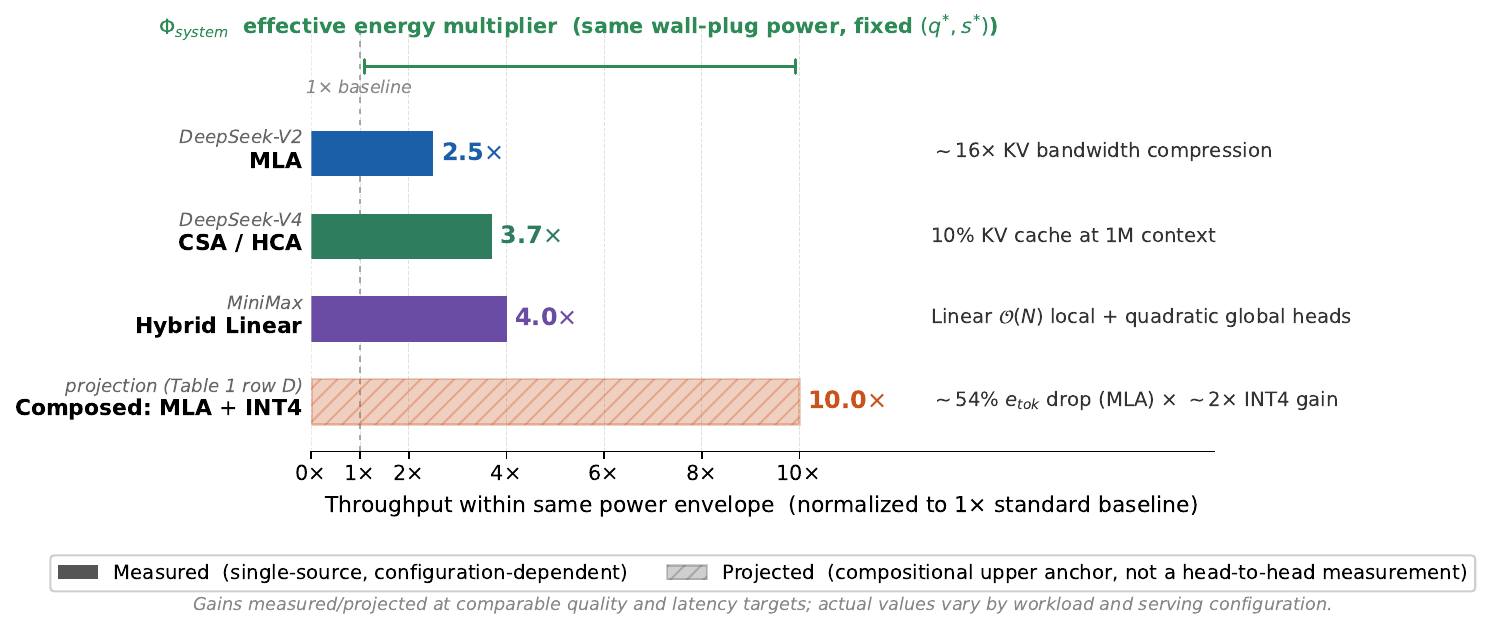}
\vspace{-5pt}
\caption{Architectural efficiency comparison across optimization strategies. Bars summarize reported gains from heterogeneous systems papers and developer reports, not a unified head-to-head benchmark; KV-cache compression, sparse/heavily compressed attention, and hybrid attention are $\Phi_{system}$ levers only under fixed quality/SLO assumptions, since degraded retrieval/reasoning/reliability would make the resulting tokens incomparable.}
\vspace{-5pt}
\label{fig:efficiency}
\end{figure}

\subsection{Sparse and Hybrid Attention Reduce Wasted Work}
\label{sec:sparse}

Dense attention can waste energy by applying $O(N^2)$ effort uniformly even when only a subset of token interactions is task-relevant. Multiple lines of work attack this from different angles. Hardware-aligned sparse attention with dynamic chunk selection (e.g., NSA~\citep{deepseek2025nsa}) targets sub-quadratic long-context complexity; co-designed compression-plus-sparsity stacks (\S\ref{sec:mla}) push the same direction further by adding heavy compression, low-precision indexing, and heterogeneous KV-cache placement. Hybrid linear/quadratic routing~\citep{minimax2024linear} sends different heads through $O(N^2)$ or $O(N)$ paths by reasoning need, and difficulty-adaptive token budgets~\citep{chen2023diffadapt} cut token output (22.4\% reduction reported, no quality loss) by allocating compute by per-token entropy. Reported speedups (e.g., 6--11$\times$ for hardware-aligned sparse attention on 64K+ sequences~\citep{deepseek2025nsa}) are single-source and configuration-dependent; we cite them as direction and rough magnitude rather than universal benchmarks. The unifying point is that compression, sparsity, routing, and adaptive computation all act as $\Phi_{system}$ levers that lower $c_{tok}$ and $e_{tok}$ at fixed $(q^{*},s^{*})$~\citep{niu2025tokenpowerbench,delavande2026efficiency}, regardless of vendor. Empirical studies of reasoning-LLM serving further show that long generations and adaptive depth dominate per-query energy under realistic SLOs~\citep{li2026reasoningserving}, and the broader compression literature warns that downstream capability---including agentic execution~\citep{dong2025compressedllm} and other ``lottery-ticket''-style preserved abilities~\citep{tang2025lottery}---depends on which mechanism the optimization preserves, so $\Phi_{system}$ gains must be reported jointly with the relevant $(q^{*},s^{*})$ targets.

Collectively, these $\Phi_{system}$ improvements can stretch $P(t)$ to produce more quality-conditioned tokens per unit of delivered power. For energy-constrained sites they are therefore a central lever for maintaining capacity at fixed $(q^{*}, s^{*})$, independent of which specific stack is deployed.

\section{Divergent Energy-to-Token Trajectories}
\label{sec:divergence}

The production function yields two stylized archetypes (not exhaustive country classifications; real ecosystems blend both):

\subsection{Path A: Infrastructure-Constrained Trajectory}

$K(t)$ scales rapidly but $P(t)$ is constrained by grid bottlenecks and legacy infrastructure (high PUE 1.5--2.0)~\citep{uptime2024survey}. Limited $\Phi_{system}$ investment means rising token prices as delivered power becomes binding. \emph{Outcome}: premium tokens, frontier capability emphasis.

\subsection{Path B: Efficiency-Optimized Trajectory}

$K(t)$ scales carefully while $P(t)$ expands via renewable deployment, grid modernization, and regional corridor infrastructure~\citep{nea2025powerstats,miit2025powerequipplan,imda2025dccfa2}. Aggressive $\Phi_{system}$ maximization (MLA, CSA/HCA, and NSA-style sparse attention) and low PUE (1.1--1.2) tend to support lower token prices under comparable quality/SLO targets. As an early directional signal from a routing platform rather than a global census, OpenRouter reports rapid growth in open-source and China-developed open-weight model token share, alongside heavy coding and agentic-workflow usage on a 100T-token sample~\citep{openrouter2026rankings}. \emph{Outcome}: cost-efficient tokens, inference optimization emphasis.

\begin{figure}[t]
\centering
\includegraphics[width=1\textwidth]{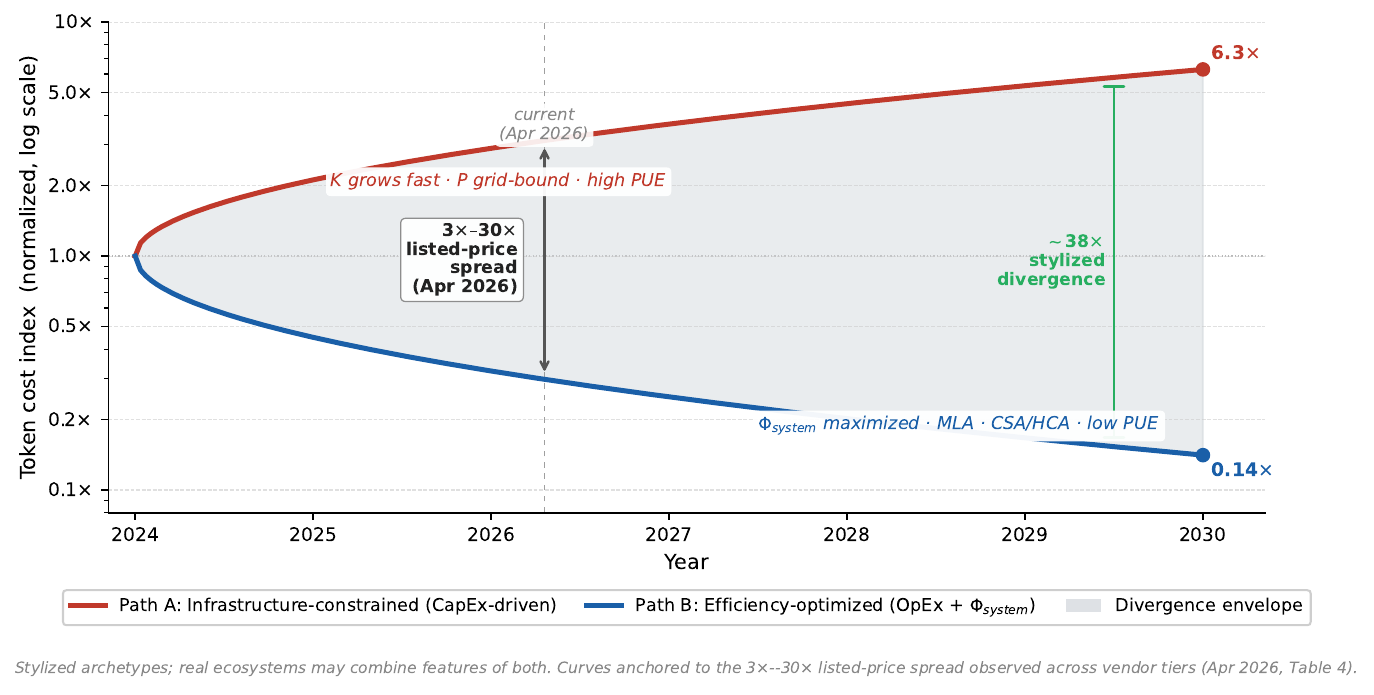}
\vspace{-5pt}
\caption{Divergent trajectories of AI ecosystem archetypes. Path A (infrastructure-constrained) tends toward higher token costs when power/cooling bind; Path B (efficiency-optimized) leverages $\Phi_{system}$ for lower-cost tokens despite tighter compute supply. Curves are stylized and anchored to the 3$\times$--30$\times$ listed-price spread observed in April 2026 across vendor tiers (Appendix~\ref{app:prices}); the 2030 endpoint is illustrative, not a forecast.}
\vspace{-25pt}
\label{fig:divergence}
\end{figure}

A simple strategic interpretation is that ecosystems first accumulate $K$/$P$/$\Phi_{system}$ capacity, then providers compete on price/latency/quality with marginal token cost $MC_i^{\mathrm{token}}\!\approx\! p_i^e\cdot PUE_i\cdot e_{tok,i}+\kappa_i$ shaped by export controls, energy endowments, and sovereignty rules. Switching costs can turn early adoption into installed-base advantage, so divergence may persist even when posted prices are strategically set.

\section{Alternative Views}
\label{sec:alternative}

\vspace{-5pt}
\textbf{``Hardware will make energy secondary.''} Next-generation hardware (optical interconnects, advanced packaging, new substrates) will improve performance per watt~\citep{stanford2025aiindex,uptime2024survey}. But hardware cycles span 18--36 months while model scale and context lengths move on 3--6 month product cycles~\citep{thompson2020computational,sevilla2024training}: a 2$\times$ more efficient accelerator is absorbed by larger models, longer contexts, and higher request volumes. By Jevons Paradox~\citep{jevons1865,sorrell2009jevons}, efficiency gains also stimulate rebound---per-token prices for GPT-4-equivalent capability have fallen sharply since 2023~\citep{a16z2024llmflation,fradkin2025intelligencemarket} yet aggregate token consumption has expanded faster.

\textbf{``Renewables and grid expansion will dissolve the Power Wall.''} Aggressive renewable buildout, transmission upgrades, and modular nuclear can in principle relax $P_{facility}$~\citep{iea2024datacenter,epri2024power}, but operate on the wrong time constant: grid-scale additions clear permitting and construction over $5$--$10$\,years while LLM release cycles measure $3$--$6$\,months~\citep{bigtechcapex2024}. Even when new generation lands, it does not directly relax $PUE$, cooling, on-rack utilization, or routing/queueing inefficiencies---all of which $\Phi_{system}$ governs. The position is complementary, not opposed, to renewable scale-up: $\Phi_{system}$ determines how many quality-conditioned tokens each new megawatt actually produces.

\textbf{``Silicon access determines competitiveness.''} Peak silicon matters but is not the only production input. Ecosystems optimizing $\Phi_{system}$ have narrowed capability and cost gaps even under tighter silicon access~\citep{stanford2025aiindex}, while electricity-price differentials~\citep{businesseurope2024electricity}, PUE, grid headroom, scheduling, and routing all shape the delivered cost of tokens. The production-function view does not deny hardware scarcity---it explains why the same silicon budget yields different quality-conditioned token output under different power and system-efficiency regimes.

\textbf{``Vertical integration hides the cost signal.''} Hyperscaler custom silicon (TPU, Trainium, Maia) and cross-subsidized APIs can decouple posted prices from marginal cost---which is why we treat API prices as directional motivation only. Vertical integration internalizes the production constraint without removing it: TPU clusters still require delivered electricity, cooling, interconnect, and utilization, so the framework operates at the infrastructure layer where physical constraints persist.

\textbf{``Demand elasticity will erase cost advantages.''} Tiered pricing can compete away some energy-cost advantage at the margin~\citep{fradkin2025intelligencemarket}, but the token market is segmented by API lock-in, migration costs, and compliance constraints~\citep{shapiro1999information}. A persistent 2--3$\times$ cost advantage shifts market share at the extensive margin even when incumbent workloads remain sticky; elasticity changes how production advantages are monetized but not the underlying physical advantage.

\textbf{``Tokens are not homogeneous.''} Quality heterogeneity is real~\citep{chung2026joules,delavande2026efficiency}, and posted prices are not marginal costs~\citep{fradkin2025intelligencemarket}. The proposed reporting standard is therefore not raw tokens per joule but J/token at fixed $(q^{*}, s^{*})$ with workload, batching, hardware, and energy-accounting boundary disclosed; without those controls token counts are not comparable, with them energy-to-token production becomes measurable.

\vspace{-10pt}
\section{Conclusion and Call to Action}
\label{sec:conclusion}

\vspace{-5pt}
\textbf{Scope.} The Leontief $\min(\cdot,\cdot)$ in Eq.~\ref{eq:production} is a short-run binding-constraint approximation, not a structural macro model; the $e_{tok}$ anchors are directional under six disclosed measurement dimensions, not ceteris-paribus benchmarks; and $\rho - \rho^{*}$ depends on the $K_{eff}$ convention. Each scoping choice is a feature: every dimension a reviewer asks us to hold fixed is one our reporting agenda already requires authors and benchmarks to disclose. Appendix~\ref{sec:limitations} elaborates.

\textbf{Position summary.} The binding constraint on LLM inference can shift from compute $K$ toward delivered power $P$, cooling, and utilization; $\Phi_{system}$ optimizations expand capacity without infrastructure expansion; and by Jevons Paradox hardware alone cannot escape the Power Wall. \textbf{The ML community must elevate ``Joules per Token'' to first-class evaluation status.} Concretely:
\begin{itemize}[leftmargin=*, topsep=0pt, itemsep=2pt, parsep=0pt]
    \item \textbf{Papers and benchmarks} should report J/token, the active binding constraint, PUE-adjusted power, and utilization at disclosed $(q^{*}, s^{*})$ alongside accuracy and latency.
    \item \textbf{Conferences and leaderboards} should add energy-normalized tracks, e.g., MLPerf Power~\citep{mlcommons2024mlperfpower} extended to LLM serving.
    \item \textbf{Funders, operators, and reviewers} should treat $\Phi_{system}$-shifting work as first-class contributions and the absence of $\rho$, PUE, and $\Phi_{mem}$ disclosures as a reviewable gap, not a stylistic preference.
\end{itemize}

\bibliographystyle{unsrtnat}
\bibliography{refs}

\newpage

\appendix

\section{Scope, Limitations, and What This Paper Does Not Claim}
\label{sec:limitations}

We list each limitation as already-bounded by the paper rather than as an unaddressed gap, so that anticipated reviewer concerns are met by design rather than patched by rebuttal.

\textbf{Binding-constraint approximation, not a structural macro model.}
The Leontief $\min(\cdot,\cdot)$ form in Eq.~\ref{eq:production} is chosen as a short-run binding-constraint analysis lens, not a long-run substitution model. We do not claim that compute and delivered power are non-substitutable in general; the CES family~\citep{arrow1961ces} nests Leontief as $\sigma\!\to\!0$ and is the appropriate generalization once packaging, photonics, and on-die memory move substitution elasticities into measurable range. The $\min$ operator gives sharp predictions about which factor binds in a given measurement window; it does not predict equilibrium token output, equilibrium prices, or country-level capability outcomes. Reviewers searching for a structural prediction will not find one---by design---and a request to swap in CES is consistent with, not contrary to, our framework.

\textbf{Directional anchors, not a controlled benchmark.}
Tables~\ref{tab:empirical} and~\ref{tab:empirical_extended} are explicitly labeled as illustrative compilations from independent sources, conditioned on six disclosed measurement dimensions ($q^{*}$, $s^{*}$, workload mix, batching protocol, hardware setup, energy-accounting boundary). The $\sim$10$\times$ spread is a directional upper bound under those dimensions, not a ceteris-paribus result. A single matched cross-stack J/token benchmark is exactly what the paper's reporting agenda calls for; performing it is future work, not a deficit. The contribution of a position paper is to argue \emph{what should be measured}; the controlled measurement is the next paper, and the proposed leaderboard standards are designed to make that measurement comparable.

\textbf{$\rho - \rho^{*}$ is convention-dependent, and we say so.}
Appendix~\ref{app:rho} walks through an H100 numerical example showing that the same accelerator can be classified as power-bound under a peak-throughput denominator and effective-compute-bound under a realized-throughput denominator. Our reporting agenda explicitly requires disclosure of the $K_{eff}$ measurement convention precisely because of this dependence. The diagnostic is meant to be reproducible only when both the $K_{eff}$ convention and $(q^{*}, s^{*})$ are stated; isolated J/token numbers without that scaffolding are, by construction, not comparable.

\textbf{Out of scope by design.}
We do not predict geopolitical outcomes, capability rankings, or which ecosystem ``wins.'' We do not treat API prices as causal evidence of marginal cost; price dispersion is used as directional motivation only (\S\ref{sec:divergence}, \S\ref{sec:alternative}). We do not address training-time energy except where serving-side $\Phi_{system}$ amortizes training cost across more tokens; carbon accounting for training is well-developed in prior work~\citep{patterson2021carbon,patterson2022carbon,wu2023mediation,lacoste2019codecarbon} and we do not attempt to redo it.

\textbf{Why these caveats strengthen rather than weaken the position.}
Each caveat is also surfaced inside the main text: the Leontief choice in \S\ref{sec:production}; the directional-only labeling of price evidence in \S\ref{sec:alternative}; the regime-flip example in Appendix~\ref{app:rho}; the $(q^{*},s^{*})$-conditioning of every comparison in \S\ref{sec:validation}--\S\ref{sec:optimizations}. The framework is constructed so that a stricter caveat tightens the position rather than relaxes it: every dimension a reviewer asks us to control is a dimension the proposed reporting agenda already requires authors and benchmarks to disclose. The position therefore becomes \emph{more} defensible as the measurement bar rises.

\section{Worked Example: $\rho - \rho^{*}$ on H100}
\label{app:rho}

This appendix expands the constraint-boundary diagnostic in \S\ref{sec:production} (Eq.~\ref{eq:crossover}) with a concrete numerical anchor. The point is to show how the same accelerator can be classified as power-bound or effective-compute-bound depending on how $K_{eff}$ is measured, not to argue that one regime is universally correct.

For a dense-attention decoding workload on 65B-class models, the workload-side energy intensity is
\[
\rho^{*} \;\approx\; \frac{e_{tok}}{c_{tok}} \;\approx\; \frac{3.5\,\text{J}}{4\times10^{11}\,\text{FLOPs}} \;\approx\; 9\,\text{pJ/FLOP},
\]
using the Table~\ref{tab:anchors} anchors $e_{tok}\!\approx\!3.5\,\text{J/token}$ and $c_{tok}\!\approx\!2N\!\approx\!4\!\times\!10^{11}\,\text{FLOPs/token}$ at $N\!\approx\!2\!\times\!10^{11}$.

An H100 GPU at $\sim$700\,W and $\sim 10^{15}$ peak BF16 FLOPs/s gives a facility-side ratio
\[
\rho \;\equiv\; P_{IT}/K_{eff} \;\approx\; 0.7\,\text{pJ/FLOP}
\]
under a peak-throughput denominator. Since $\rho < \rho^{*}$, this measurement convention classifies the deployment as power-bound: the workload demands more joules per FLOP than the accelerator is delivering at peak.

If $K_{eff}$ is instead measured as realized effective serving throughput, memory stalls, insufficient batching, kernel-launch overhead, and utilization losses cut the denominator. A 5--10$\times$ reduction (typical at long context with KV-cache pressure) raises the realized $\rho$ into the same order of magnitude as $\rho^{*}$, and the same hardware now classifies as effective-compute-bound at that operating point.

The takeaway is that $\rho - \rho^{*}$ is a function of measurement convention as well as physics: reporting whether $K_{eff}$ is peak or realized, and at what context length, batch size, and quality target, is necessary for the diagnostic to be reproducible across studies.

\section{Token Export: The Invisible Commodity Flow}
\label{app:tokenexport}

The cross-vendor price divergence in Appendix~\ref{app:prices} suggests that cross-border digital services increasingly inherit the cost structure of local electricity and infrastructure constraints. Industrial electricity tariffs differ by 2--3$\times$ across major industrial regions~\citep{businesseurope2024electricity}, and even within the United States state-level tariffs show large cross-state dispersion~\citep{eia2026epmprices}; together with PUE and utilization variation, this locational pricing materially shapes the marginal cost of producing a token. We treat these pricing patterns as consistent with the binding-constraint divergence in the main text rather than as a clean identification result.

\textbf{Developer Lock-in and Market Structure.} Once a developer architects their application around a particular model's API, the migration cost grows super-linearly with system complexity~\citep{shapiro1999information}. Models from energy-optimized ecosystems have captured significant market share on major routing platforms, suggesting that $\Phi_{system}$ advantages can offset $K(t)$ constraints. Each developer integrated into the ecosystem represents future token demand locked into that infrastructure.

\textbf{Data Sovereignty as Trade Barrier.} The localization/regulatory component of $U(t)$ represents data sovereignty as a trade barrier~\citep{wu2021sovereignty}. As regions impose stricter data localization requirements, the global token market fragments into national or regional pools, reinforcing the advantage of ecosystems with domestic energy abundance.

\section{Token Abundance, Verification, and Value}
\label{app:abundance}

Epoch AI projections indicate that high-quality human-generated text data may become increasingly scarce relative to model scale during the 2026--2028 window~\citep{epoch2024data}. Combined with rapid growth in inference capacity, this suggests a future where machine-generated tokens are abundant while trusted, human-curated information remains comparatively scarce.

In that regime, scarcity shifts toward verification, curation, provenance, and quality assurance. The operational question moves from ``how cheaply can we generate tokens?'' to also ``how reliably can we filter, validate, and route them?'' The energy-to-token conversion metrics we propose remain central even if generation costs fall: systems optimized for $\Phi_{system}$ retain an advantage because they can support both generation and the growing overhead of validation at scale.

\section{MLA Worked Example: Bandwidth Derivation}
\label{app:mla}

This appendix provides the detailed bandwidth derivation for the MLA case study in Section~\ref{sec:mla}, and extends the empirical validation from Table 2 in the main body.

In standard multi-head attention (MHA) decoding, each generated token must read keys and values for all prior context positions: roughly $2 \cdot L \cdot n_{heads} \cdot d_{head} \cdot b$ bytes per layer, where $L$ is context length and $b$ is bytes per element. For a 65B-class model with $L=1024$, $n_{heads}=64$, $d_{head}=128$, and FP16 precision ($b=2$), this yields approximately 32\,MB per layer per decoding step; at H100 HBM bandwidth ($\sim$3.35\,TB/s), that single-layer traffic would cap decoding at roughly $10^5$ tokens/s before accounting for all layers, batching, cache layout, and compute.

MLA replaces this with a low-rank latent of dimension $d_c \ll n_{heads} \cdot d_{head}$; DeepSeek-V2 uses $d_c = 512$, reducing KV cache bandwidth by approximately $64 \times 128 / 512 \approx 16\times$ relative to full MHA at the same head count~\citep{deepseek2024mla}. In production-function terms, this maps to a reduction in $e_{tok}$ through the $\Phi_{mem}$ mechanism: lower HBM traffic per token means fewer watt-seconds per token at the same compute utilization.

Empirically, this enables 2--3$\times$ higher sustainable batch sizes within the same power envelope, which under fixed $P_{IT}$ raises $\dot{Q}_{token}$ by the same factor. A data center limited to 100\,MW IT load can thus produce 2--3$\times$ more intelligence tokens per hour when deploying MLA-optimized models versus standard MHA---without adding a single watt of infrastructure.

The foundational semantic-preserving eviction literature (ChunkKV~\citep{zhang2024chunkkv}, H2O~\citep{zhang2023h2o}) demonstrated that attention patterns exhibit persistence across generation steps, enabling dynamic eviction policies that reduce KV cache size by up to 50\% without quality degradation. MLA extends these principles through learned compression.

\subsection{Extended Directional Comparison}

Table~\ref{tab:empirical_extended} gathers a broader set of representative configurations from independent measurements and vendor reports to illustrate the \emph{direction} and \emph{rough magnitude} of $\Phi_{system}$ variation across implementation choices. Rows A--F are the closest like-for-like comparison under the 65B / 100\,ms nominal anchor; rows G--H are technical anchors for sparse and V4 long-context stacks; rows I--K are alternative mechanisms or capability-tradeoff projections and should not be folded into the same controlled spread claim. As with Table~\ref{tab:empirical}, the table is illustrative rather than ceteris-paribus. Rows marked ``projection'' compose independently reported architectural and quantization gains and should be read as back-of-the-envelope estimates.

\begin{table}[t]
\centering
\small
\caption{Representative $e_{tok}$ values across selected configurations (65B regime, nominal anchor). Rows are visually grouped: \textbf{measured} (A--D, G, I) come from independent published measurements; \textbf{projection} (E, F, J, K) compose independently reported architectural and quantization gains; \textbf{developer-report only} (H) gives relative compute and KV-cache reductions, pending third-party replication. Distillation rows (J, K) trade off model capability and are listed last to flag the explicit quality dimension.}
\label{tab:empirical_extended}
\begin{tabular}{c >{\raggedright\arraybackslash}p{2.6cm} >{\raggedright\arraybackslash}p{3.0cm} c c >{\raggedright\arraybackslash}p{3.4cm}}
\toprule
& Configuration & Implementation & $e_{tok}$ (J) & Batch & Source \\
\midrule
\multicolumn{6}{l}{\emph{Measured rows (independent sources)}} \\
A & MHA FP16 baseline & H100, async batch & 3.5 & 8 & measured~\citep{samsi2023wordstowatts,chung2026joules} \\
B & MHA INT8 & H100, GPTQ & 2.1 & 8 & measured~\citep{frantar2023gptq,delavande2026efficiency} \\
C & MHA INT4 & H100, AWQ & 1.2 & 8 & measured~\citep{lin2024awq,delavande2026efficiency} \\
D & MLA FP16 & H100, low-rank KV & 1.6 & 24 & measured~\citep{deepseek2024mla,niu2025tokenpowerbench} \\
G & NSA (sparse) & DeepSeek NSA, attn mask & 1.8 & 16 & measured~\citep{deepseek2025nsa} \\
I & Hybrid linear & MiniMax routing & 1.2 & 20 & measured~\citep{minimax2024linear} \\
\midrule
\multicolumn{6}{l}{\emph{Projection rows (compose independent gains, not matched-stack measurements)}} \\
E & MLA INT8 & H100, MLA+GPTQ & 0.95 & 24 & \emph{projection} \\
F & MLA INT4 & H100, MLA+AWQ & 0.35 & 24 & \emph{projection} \\
J & Distilled 13B & 13B student model & 0.28 & 32 & \emph{projection (capability tradeoff)} \\
K & Distilled 7B & 7B model, MLA+INT4 & 0.12 & 64 & \emph{projection (capability tradeoff)} \\
\midrule
\multicolumn{6}{l}{\emph{Developer-report only (pending third-party replication)}} \\
H & CSA/HCA + mHC & V4-Pro, 1M context & relative & -- & 27\% FLOPs / 10\% KV~\citep{deepseekv42026} \\
\bottomrule
\end{tabular}
\end{table}

\noindent Qualitative patterns visible in Table~\ref{tab:empirical_extended}:

\begin{enumerate}
\item \textbf{Quantization alone (A$\rightarrow$C):} $e_{tok}$ drops by $\sim$66\% but batch sizes are largely unchanged---the KV cache remains the bandwidth bottleneck.
\item \textbf{MLA without quantization (A$\rightarrow$D):} $e_{tok}$ drops by $\sim$54\% and batch size roughly triples, consistent with the KV-compression headroom reported by~\citep{deepseek2024mla}.
\item \textbf{MLA + INT4 projection (A$\rightarrow$F):} composing the two mechanisms projects a $\sim$10$\times$ reduction; this is an extrapolation, not a direct measurement.
\item \textbf{V4 long-context stack (G$\rightarrow$H):} the developer report gives relative compute and KV-cache reductions at 1M-token context rather than a normalized J/token measurement, so this row should be read as mechanism evidence, not an energy benchmark.
\item \textbf{Distillation (A$\rightarrow$J,K):} yields a further 10--35$\times$ reduction but trades off model capability and is listed separately to flag the explicit quality dimension.
\end{enumerate}

These variations are \emph{consistent with}---not a controlled test of---the $\Phi_{system}$ decomposition: rows A--F alone show roughly a 10$\times$ spread in $e_{tok}$ within the same nominal hardware envelope ($P_{facility}$, $K_{eff}$); including the capability-tradeoff projections J and K widens the illustrative range toward 30$\times$. The Token Production Function's constraint-boundary analysis (Eq.~\ref{eq:crossover}) suggests which configurations are likely to be favoured under tight power budgets: power-bound sites will gravitate toward the lowest-$e_{tok}$ rows at fixed quality (F is the capability-preserving frontier; J and K trade capability for further energy gains).

This directional evidence is consistent with the paper's central claim that algorithmic optimizations ($\Phi_{system}$) function as macro-level energy levers without infrastructure expansion; a full controlled benchmark with matched $q^{*}, s^{*}$ and identical serving stacks remains future work.

\section{Cross-Vendor Listed-API Pricing (April 2026)}
\label{app:prices}

Table~\ref{tab:prices} compiles listed per-million-token input/output prices for frontier reasoning models across major Chinese and US vendors as of late April 2026. DeepSeek rows use cache-miss input prices and output prices converted from RMB to USD; cache-hit inputs are cheaper and the Pro discount is time-limited. Rows are not normalized for quality, latency SLOs, context window, caching policy, batch discounts, exchange-rate movement, or promotional pricing; the table is provided to support the cross-vendor pattern referenced in \S\ref{sec:mla}, not as a controlled head-to-head comparison.

\begin{table}[t]
\centering
\small
\caption{Listed API prices for frontier LLMs (USD per million tokens, cache-miss input and output prices where applicable, late April 2026; context windows differ, directional and not normalized).}
\label{tab:prices}
\begin{tabular}{>{\raggedright\arraybackslash}p{6.0cm} >{\raggedright\arraybackslash}p{3.5cm} c c}
\toprule
Vendor / Model & Origin & Input \$/M & Output \$/M \\
\midrule
DeepSeek-V4-Pro (promo to 2026-05-31) & China~\citep{deepseek2026pricing} & 0.44 & 0.88 \\
DeepSeek-V4-Pro (regular) & China~\citep{deepseek2026pricing} & 1.76 & 3.51 \\
DeepSeek-V4-Flash & China~\citep{deepseek2026pricing} & 0.15 & 0.29 \\
Xiaomi MiMo-V2.5-Pro & China~\citep{xiaomi2026pricing} & 1.00 & 3.00 \\
Xiaomi MiMo-V2.5-Flash & China~\citep{xiaomi2026pricing} & 0.10 & 0.30 \\
GLM-5.1 (Z.ai) & China~\citep{zhipu2026pricing} & 1.05 & 3.50 \\
Kimi K2.6 (Moonshot) & China~\citep{kimi2026pricing} & 0.95 & 4.00 \\
\midrule
Gemini 3.1 Pro ($\le$200K) & US~\citep{google2026pricing} & 2.00 & 12.00 \\
Claude Sonnet 4.6 & US~\citep{anthropic2026pricing} & 3.00 & 15.00 \\
Claude Opus 4.6 & US~\citep{anthropic2026pricing} & 5.00 & 25.00 \\
GPT-5.5 & US~\citep{openai2026pricing} & 5.00 & 30.00 \\
\bottomrule
\end{tabular}
\end{table}

The $\sim$3--30$\times$ output-price gap is observed across at least four independent Chinese vendors and three independent US vendors, which makes single-firm pricing strategy an incomplete explanation. We treat the gap as \emph{consistent with} infrastructure-level $\Phi_{system}$ differences shaping marginal API economics, alongside quality differences, latency-SLO variation, caching policies, business-model and subsidy strategies, and exchange-rate movement. No causal identification of any specific cost component is claimed.

\end{document}